\documentstyle[epsfig]{aipproc}
\begin{document}
\title{Short Time-scale Gamma-Ray Variability of Blazars and EGRET
Unidentified Sources}

\author{S.D. Bloom$^1$, D.J. Thompson$^2$, R. C. Hartman$^2$,\\
C. von Montigny$^3$}
\address{$^1$NASA/GSFC. NAS/NRC Resident Research Associate\\
$^2$NASA/GSFC.\\
$^3$Landessternwarte Heidelberg}

\maketitle

\begin{abstract}

We have begun to examine the EGRET database for short term variations
in the fluxes of blazars and the unidentified sources at high Galactic
latitudes. 
We find that several AGN show previously unreported 
variability. Such variations are consistent with 
inverse Compton scattering processes in a shock propagating through a 
relativistic jet.   

\end{abstract}

\section*{Introduction}

Previous gamma-ray variability studies of EGRET
detected blazars and unidentified sources have focused on long time-scale
(weeks to years)
variability \cite{sdbmuk97},\cite{sdbmcl96}. However, observations of bright flares, such as
the outbursts of PKS 1622-297 \cite{sdbmat97a} and 3C 279 (\cite{sdbhar96},
\cite{sdbweh97}), clearly show variabilty on 
time-scales much less than a week. We therefore have set out to examine
the EGRET data for variations on short time-scales (roughly,
2--14 days).

In general, observations of at least two weeks had been scheduled for
most EGRET projects in order to detect weak sources.
The standard counts maps were generated for the entire viewing period, and thus
fluxes and spectral information determined from them are {\it averaged} properties over many days. Thus, any shorter time-scale variations would have been
washed out. Since several blazars are known to vary dramatically on time-scales $\sim$
1 day--1 week
at gamma-ray energies , we would expect that there are other blazars with
similar, though probably less dramatic, variability properties. We have thus started an examination of
EGRET data for such fluctuations.
This process will also bring our attention to bright
transients that only appear in part of one viewing period. 

\section*{Method}
In order to concentrate on blazars and potential blazars
we have limited our study to viewing periods (VP's) centered on 
$b > \vert 20^\circ \vert$. A similar study on the Galactic anticenter
region is reported on by Thompson {\it et al} \cite{sdbtho97}. 
Similar to Thompson {\it et al} \cite{sdbtho97}, we have divided each viewing 
period into 
new two-day maps of $>100$ MeV counts (accomplished for 35 VP's).
A standard maximum likelihood analysis was performed on each map
\cite{sdbmat96} to determine fluxes over two-day scales. An estimated
systematic uncertainty of 6.5 \% has been added to the statistical
uncertainties \cite{sdbmcl96}.
We have then used a  $\chi^2$ analysis on fluxes and determined
probability, $P$, that the data are consistent with a constant flux. 
The variability parameter of Mc Laughlin {\it et al} \cite{sdbmcl96} is used to
easily compare significance of variability among sources (and different
VP's on the same source):

\[ 
V \equiv \vert log P \vert 
\]

\section*{Results }
The blazar PKS 0208-512 (Figure 1) shows a factor of 2--3 increase 
over 3 weeks, followed by a rapid (2 day) factor of 3 decrease, and then
another factor of 2--3 increase over the last week. 
The time history of the unidentified source 2EG J1835+5919 (Figure 2)
is consistent with no variability over the 30 day period.
However, inspection of Figure 2 reveals that there may have been
a drop in flux over the last 10 days. The possible
blazar 2EG J0220+4228 (0219+428) (Figure 3) shows a factor of 2--3 
increase and decrease within one week. The blazar 0446+112 (Figure 4) undergoes
a nearly factor of 4 decrease over one month, and nearly a factor of
two within the last two weeks. No transients were found
in any of the viewing periods studied.

\section*{Discussion}
The shortest time-scales of variability observed in this study 
are consistent with the model of Romanova and Lovelace \cite{sdbrom97}.
They consider inverse Compton processes within a shock moving through
a relativistic jet. A sudden acceleration of particles causes a
brief synchrotron flare ($<< 1$ day) which will be accompanied 
by a simultaneous synchrotron self-Compton (SSC) flare over
similar time-scales. However, since external inverse Compton models, such
as the ``mirror'' model of Ghisellini \& Madau \cite{sdbghi96}, occur over
larger spatial scales than the SSC models, the high energy flares will
be delayed from the synchrotron flare and occur over several days. 
It is beyond the scope of this study to attempt any multiwavelength
analysis and detailed theoretical modeling. In addition, we
can not time-resolve the gamma-rays over $< 1$ day; however, our future
work will check for general consistency with these models. The
relatively low amplitude and long duration of the flares in this study
(as compared to, say, 1622-297 and 3C 279) favor external scattering
models \cite{sdbrom97}.

\section*{Conclusion}
We have shown that 
three blazars undergo factors of 2--3 variability on time-scales $\sim 1$ week.
This variability is consistent with inverse Compton processes 
from a shock in a jet. The high amplitude variability seen within one
week for 2EG J0220+4228 suggests that this source is much more likely
to be related to the blazar 0219+428  than to the pulsar PSR J0218+4232
\cite{sdbver96} at energies $>100$ MeV. 2EG J1835+5919 shows only marginal 
short time-scale variability,
and varies by less than a factor of two over longer time-scales
\cite{sdbtho95}.
There are no compelling radio counterparts to this source
\cite{sdbmat97b}.
These results suggest that the source 2EG J1835+5919 is unlikely to be a 
blazar.

\begin{figure}[b!] 
\centerline{\epsfig{file=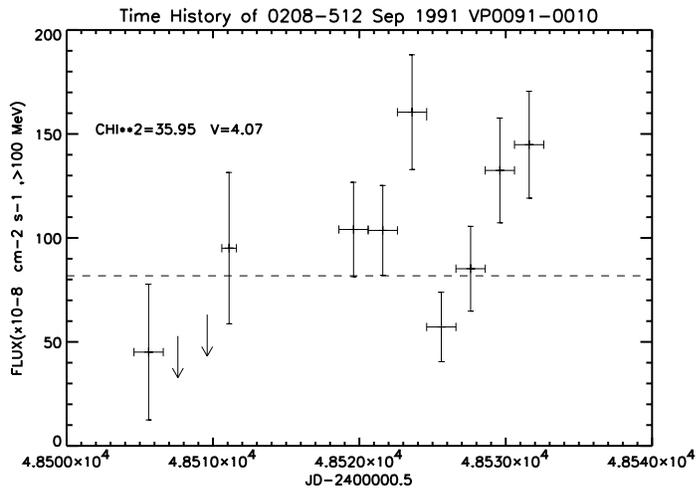,height=9cm,bbllx=50pt,bblly=240pt,bburx=640pt,bbury=730pt,clip=.}}
\vspace{10pt}
\caption{Short-term flux history for blazar PKS 0208-512 }
\label{fig1}
\end{figure}

\begin{figure}[b!] 
\centerline{\epsfig{file=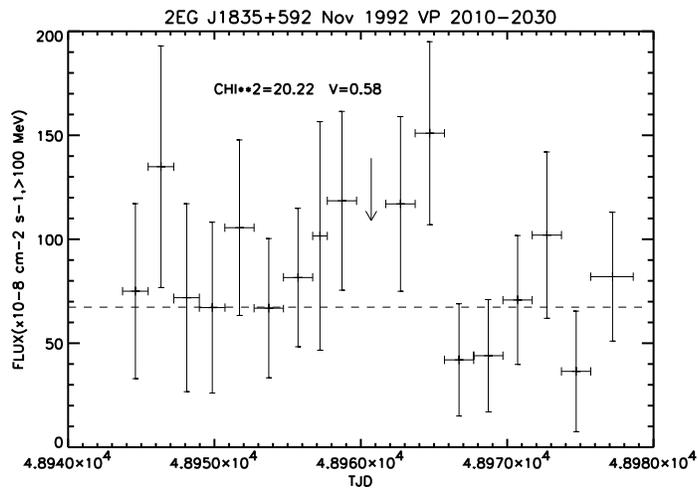,height=9cm,bbllx=60pt,bblly=240pt,bburx=680pt,bbury=730pt,clip=.}}
\vspace{10pt}
\caption{Short-term flux history for 2EG J1835+592 }
\label{fig2}
\end{figure}

\begin{figure}[b!] 
\centerline{\epsfig{file=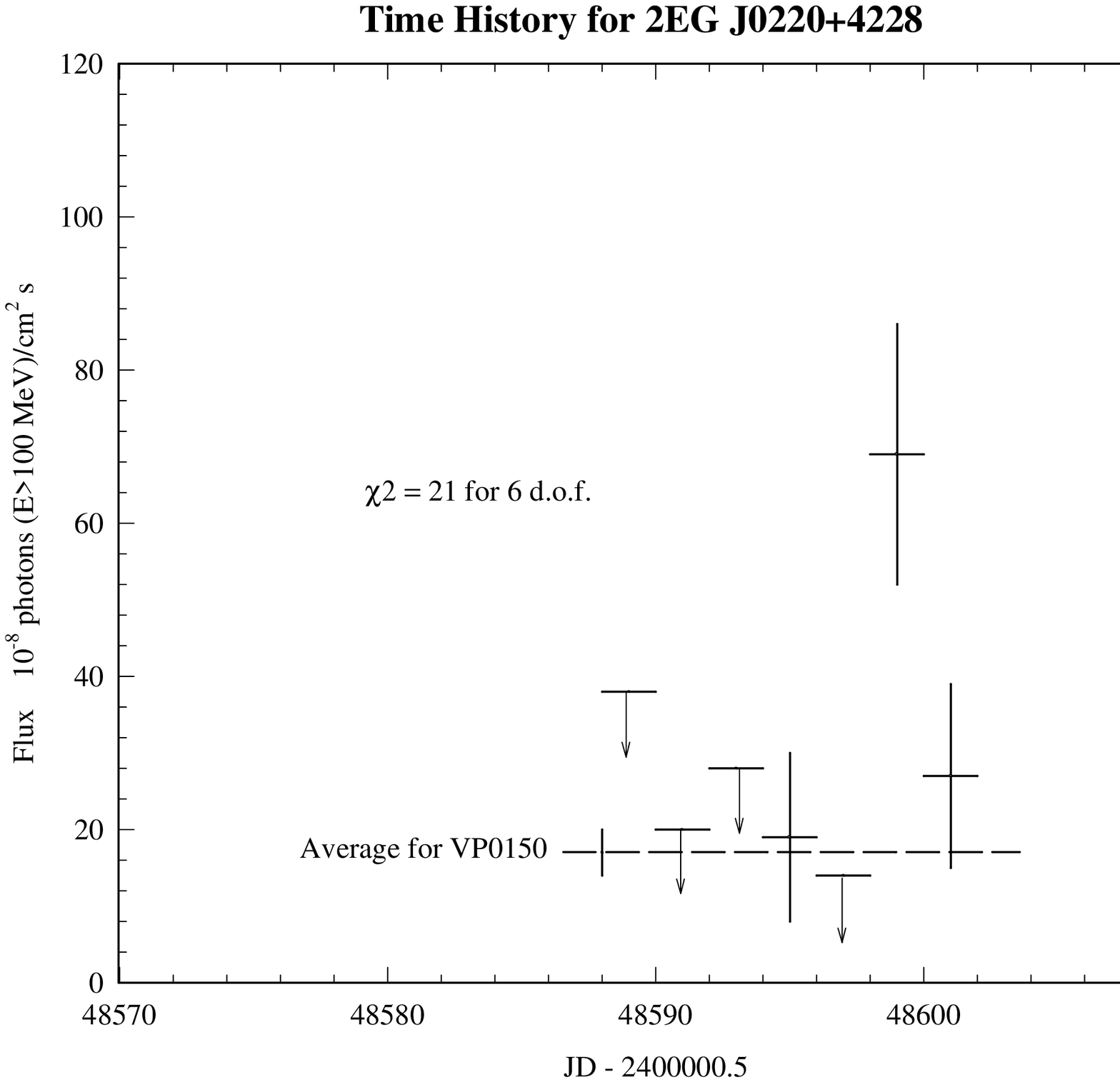,height=9cm,bbllx=50pt,bblly=240pt,bburx=800pt,bbury=730pt,clip=.}}
\vspace{10pt} 
\caption{Short-term flux history for 2EG J0220+4228. }
\label{fig3}
\end{figure}

\begin{figure}[b!] 
\centerline{\epsfig{file=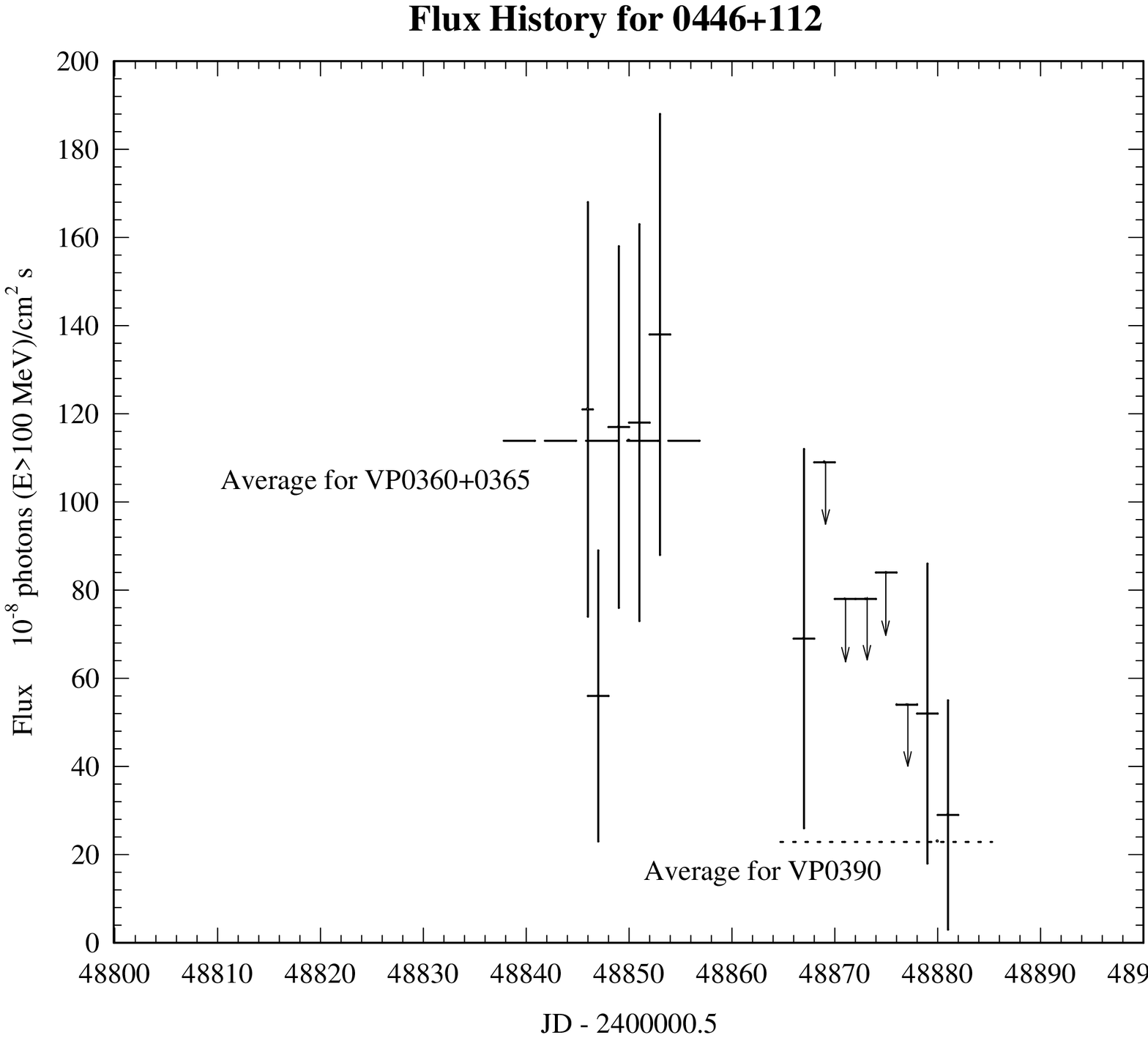,height=9cm,bbllx=50pt,bblly=240pt,bburx=640pt,bbury=730pt,clip=.}}
\vspace{10pt}
\caption{Short-term flux history for blazar 0446+112}
\label{fig4}
\end{figure}

\begin{references}
\bibitem{sdbghi96}Ghisselini, G. and Madau, P.,{\it M.N.R.A.S.},{\bf 280}, 67 (1996).
\bibitem{sdbhar96}Hartman, R. C. {\it et al}, {\it Ap. J.} {\bf 461}, 698 (1996).
\bibitem{sdbmat96}Mattox, J. R. {\it et al}, {\it Ap. J.} {\bf 461}, 39 (1996).
\bibitem{sdbmat97a}Mattox, J. R. {\it et al}, {\it Ap. J.} {\bf 476}, 692 (1997).
\bibitem{sdbmat97b}Mattox, J. R. {\it et al}, {\it Ap. J.} {\bf 481}, 95 (1997).
\bibitem{sdbmcl96}Mc Laughlin, M. A. {\it et al} , {\it Ap. J.} {\bf 473}, 763 (1996). 
\bibitem{sdbmuk97}Mukherjee, R. {\it et al}, {\it Ap. J.}, in press (1997).
\bibitem{sdbrom97}Romanova, M. M. and Lovelace, R. V. E. , {\it Ap. J.} {\bf 475}, 97 (1997).
\bibitem{sdbtho95}Thompson, D. J. {\it et al}, {\it Ap. J. Suppl.} {\bf 101}, 259 (1995).
\bibitem{sdbtho97}Thompson, D. J. {\it et al}, these proceedings (1997).
\bibitem{sdbver96}Verbunt, F. {\it et al} , {\it A.\&A.}, {\bf 311}, 9 (1996).
\bibitem{sdbweh97}Wehrle, A. E. {\it et al}, in preparation (1997).
\end{references}
\end{document}